\documentclass[5p,authoryear,twocolumn]{elsarticle}

\usepackage{graphicx}
\usepackage{amssymb}

\usepackage{hyperref}
\usepackage{breakurl}

\journal{Astronomy and Computing}

\newcommand{\planss}{Planetary and Space Science}
\newcommand{\aap}{Astronomy and Astrophysics}
\newcommand{\icarus}{Icarus}

\begin{document}

\begin{frontmatter}

\title{Asteroids@home -- A BOINC distributed computing project for asteroid shape reconstruction}

\author[au]{Josef \v{D}urech}
\author[nice]{Josef Hanu\v{s}}
\author[cnt]{Radim Van\v{c}o}

\address[au]{Astronomical Institute, Faculty of Mathematics and Physics, Charles University in Prague, V Hole\v{s}ovi\v{c}k\'ach 2, 180\,00 Prague 8, Czech Republic}
\address[nice]{UNS-CNRS-Observatoire de la C\^ote d'Azur, BP 4229, 06304 Nice Cedex 4, France}
\address[cnt]{Czech National Team}

\begin{abstract}We present the project Asteroids@home that uses distributed computing to solve the time-consuming inverse problem of shape reconstruction of asteroids. The project uses the Berkeley Open Infrastructure for Network Computing (BOINC) framework to distribute, collect, and validate small computational units that are solved independently at individual computers of volunteers connected to the project. Shapes, rotational periods, and orientations of the spin axes of asteroids are reconstructed from their disk-integrated photometry by the lightcurve inversion method. 
\end{abstract}


\end{frontmatter}

\section{Introduction}

With more than 500,000 discovered objects, asteroids form a large population of small bodies in the solar system that was affected by all processes that were acting during the formation and evolution of the solar system. By studying asteroids, we can reveal the history and current state of our cosmic neighborhood. In general, study of asteroids can be done either in situ by spacecrafts or by remote-sensing techniques. In situ measurements are limited to only a very small sample of all asteroids that are direct targets of spacecraft missions or fly-by opportunities. However, remote-sensing techniques are, in principle, feasible for most of the known population. Although the level of information we are able to get from remote sensing is inevitably much lower than from a detailed spacecraft mission, the basic physical properties can be successfully obtained for a substantial part of the asteroid population.

One of the main sources of information about asteroid shapes and spin states (i.e., rotational periods and spin axis directions) is their disk-integrated photometry that is available for all known asteroids. Because asteroids have irregular shapes and rotate, the amount of light reflected towards the observer at Earth varies with asteroid's rotation -- we observe a lightcurve. An effective method to reconstruct asteroid shapes and spin states from their lightcurves (so called lightcurve inversion) has been developed by \cite{Kaa.Tor:01, Kaa.ea:01}. 

In the following sections, we describe our project of distributed computing named Asteroids@home that is aimed for solving the lightcurve inversion problem for a substantial part of the asteroid population.

\section{Asteroid lightcurve inversion}

A method for lightcurve inversion was developed by \cite{Kaa.ea:01}. It uses all available photometric data to reconstruct a convex shape model of an asteroid (together with its sidereal rotation period and the spin axis direction) that provides the best fit to the data. The review of the method can be found in \cite{Kaa.ea:02c}. Its mathematical stability and uniqueness was proved by \cite{Lam.Kaa:01} and its results were independently confirmed by disk-resolved images \citep{Mar.ea:06,Han.ea:13a}, stellar occultation data \citep{Dur.ea:11}, or spacecraft images \citep{Kel.ea:10}. 

Although real asteroids have in general complex shapes with (sometimes large) concavities, their lightcurves can be successfully reproduced with convex shapes. Nonconvex models are an alternative to convex ones, but they lack the mathematical uniqueness and stability and in practice they are  needed only when high-quality lightcurves observed at high solar phase angles or disk-resolved data are available. In this sense, convex models should be taken as approximations to the real shapes of asteroids. 

Since the publication of the method, models of about 500 asteroids were derived by this technique \citep[][for example]{Dur.ea:09, Dur.ea:11, Han.ea:11, Han.ea:13b, Mar.ea:11}; most of them are publicly available in the Database of Asteroid Models from Inversion Techniques \citep[DAMIT\footnote{\url{http://astro.troja.mff.cuni.cz/projects/asteroids3D}},][]{Dur.ea:10}. At this site, the source codes for the lightcurve inversion called {\tt convexinv} can be downloaded. This code written in the c programming language is a default version of the shape optimization and is widely used by the scientific community. Only minor changes related to input/output file format were necessary to satisfy the BOINC specifications. Recently, the {\tt convexinv} program has been also modified to (i) deal with disk-resolved data and nonconvex features of the shape \citep{Car.ea:12}, (ii) changing rotation rate to model the YORP effect \citep{Dur.ea:08}, or (iii) excited rotation state \citep{Kaa:01, Pra.ea:14}.  

The lightcurve inversion code uses the Levenberg-Marquardt algorithm to converge to a local minimum in $\chi^2$, where $\chi^2$ is a usual measure of difference between the observed and modeled brightness $L$ taking into account the errors $\sigma$:
\begin{equation}
 \chi^2 = \sum_i \left( \frac{L_\mathrm{obs}^{(i)} - L_\mathrm{model}^{(i)}}{\sigma_i} \right)^2\,.
\end{equation}
For each epoch corresponding to the $i$-th observation, the brightness $L_\mathrm{model}^{(i)}$ is computed as a sum of contributions from surface elements that are illuminated by Sun and seen by the observer. The orientation of the model in inertial space is determined by the direction of the spin axis given in ecliptic longitude $\lambda$ and latitude $\beta$, and the rotation period $P$. The shape is represented as convex polyhedron and parametrized by spherical harmonics. The coefficients of spherical harmonics series are optimized together with the spin parameters to get the lowest value of $\chi^2$ (for more details see \cite{Kaa.ea:02c}). Depending on the resolution of the shape (degree and order of the spherical harmonics series), the number of parameters to be optimized is typically between 20 and 90.

To find the global minimum, however, we need to start at many initial points in the parameter space to go through all relevant local minima and then select the solution with the lowest $\chi^2$. Convergence of the shape towards the local minimum is robust, so the initial shape can be always a sphere, for example, while the initial spin and period parameters have to cover the whole parameter subspace. The number of initial pole directions is usually ten (isotropically distributed in ecliptic coordinates), which is enough for a safe convergence into the global minimum in the spin subspace. What makes the problem time consuming is a large number of closely packed local minima in the period subspace. The local minima are separated by about $0.5 P^2 / \Delta T$, where $P$ is the rotation period and $\Delta T$ is the length of the time interval covered by observations \citep{Kaa:01}. For a typical set of lightcurves sufficient for inversion (tens of lightcurves observed during several apparitions), the rotation period can be estimated very accurately without any modeling from the period analysis of the signal. Then the interval of periods that has to be searched for the global minimum is narrow and the process is fast. However, with the data that are sparse in time with respect to the rotation period, the classical Fourier-based or phase dispersion minimization methods cannot be used and the rotation period cannot be easily estimated from the lightcurve data. The sparse-in-time photometry is typically provided by all-sky astrometric surveys (Catalina, Pan-STARRS, Gaia satellite, for example). 
With a typical rotation period of several hours and $\sim$15 years of observations, we have to test hundreds of thousands initial periods to be sure not to miss the global minimum.

\begin{figure}[t]
\includegraphics[width=\columnwidth]{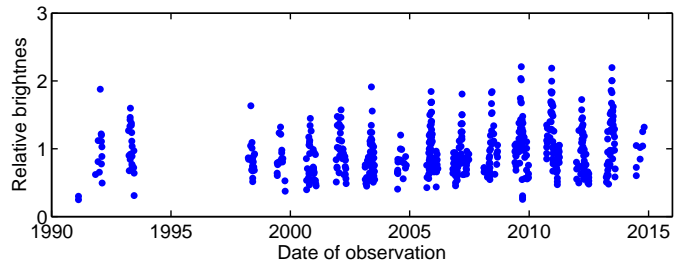}
\caption{Sparse-in-time photometry of asteroid (243)~Ida downloaded from AstDyS. There are 700 individual photometric points observed during 17 apparitions. The brightness was reduced to unit distance from Earth and Sun.}
\label{fig:Ida_data}
\end{figure}

\begin{figure}[t]
\includegraphics[width=\columnwidth]{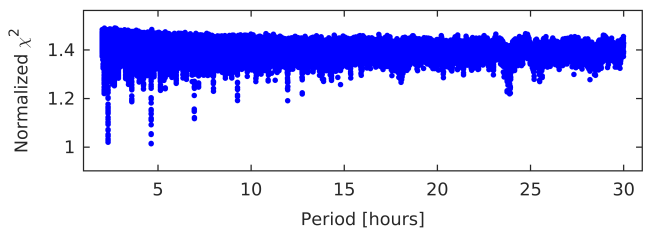}
\caption{Period search results for asteroid (243)~Ida for sparse photometry (Fig.\,\ref{fig:Ida_data}). Each point corresponds to the local minimum in the parameter space. The lowest $\chi^2$ corresponds to the best-fit model for period $P = 4.633631$\,h.}
\label{fig:Ida_per}
\end{figure}

\begin{figure}[t]
\includegraphics[width=\columnwidth]{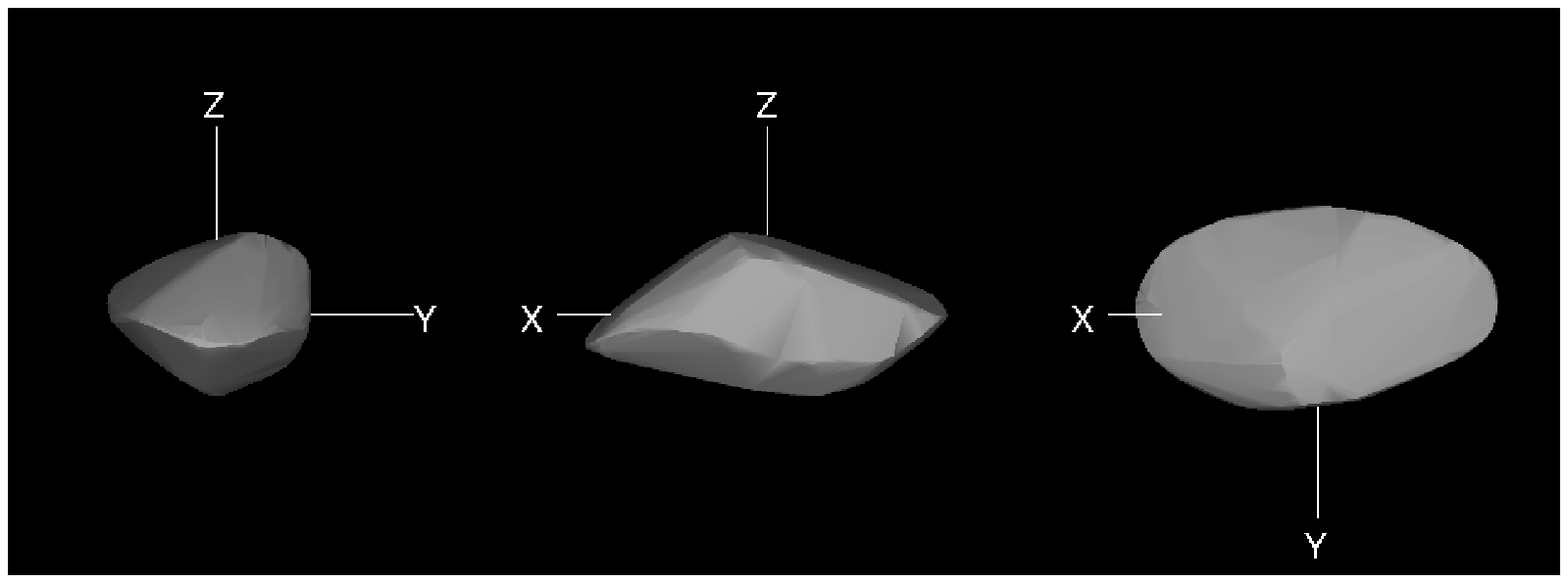}
\includegraphics[width=\columnwidth]{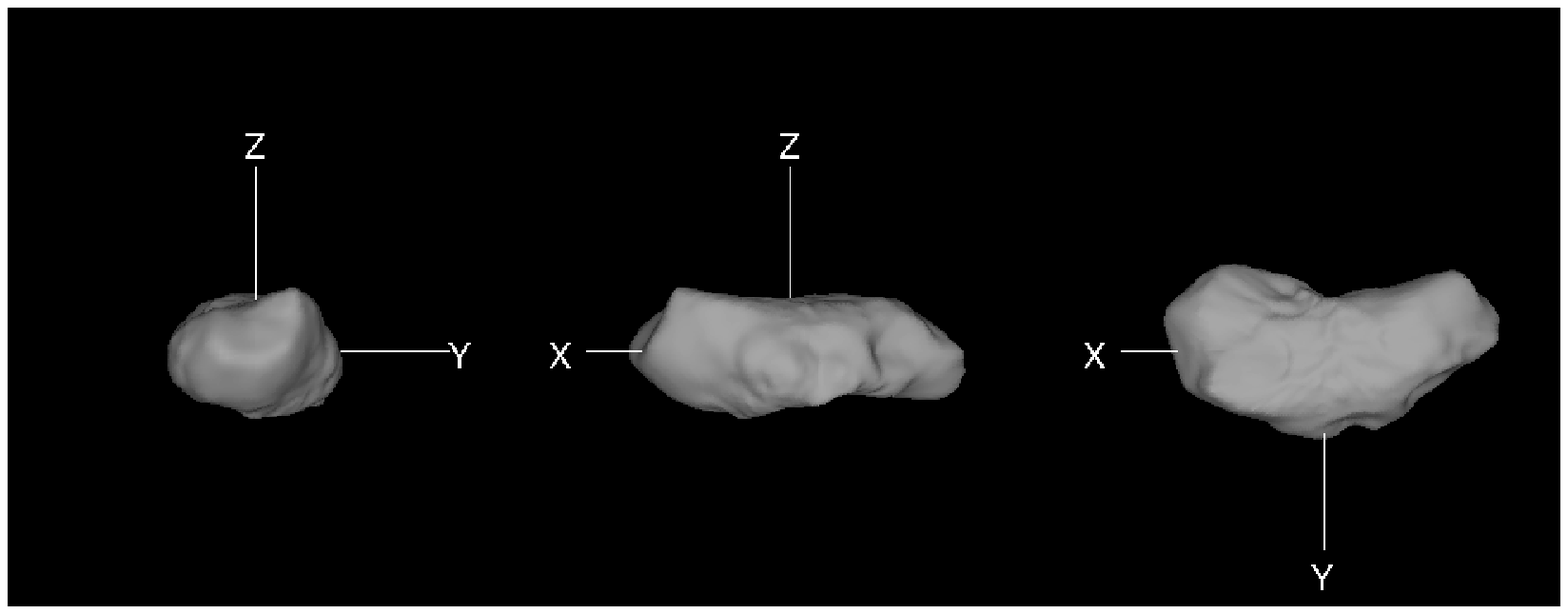}
\caption{Comparison of the convex model of (243)~Ida reconstructed from sparse photometry (top) with the detailed shape models reconstructed from fly-by images obtained by Galileo probe \citep[bottom,][]{Tho.ea:96}.}
\label{fig:Ida_shapes}
\end{figure}

As an example of sparse-in-time photometry and a corresponding model, we present results for asteroid (243)~Ida. The photometry obtained by sky surveys and downloaded from AstDyS\footnote{\url{http://hamilton.dm.unipi.it/astdys}} site is shown in Fig.~\ref{fig:Ida_data}. The brightness was reduced to the unit distance from Earth and Sun. The groups of points correspond to individual apparitions. The periodogram is shown in Fig.\,\ref{fig:Ida_per}, it consists of about 240,000 points. There are two clear minima with the lowest $\chi^2$ at 4.634 and 2.317\,h. The best-fit model corresponds to the period 4.634\,h and is shown in Fig.\,\ref{fig:Ida_shapes}. The model is compared with the real shape of Ida as reconstructed from the fly-by images obtained by Galileo probe \citep{Tho.ea:96}. The pole direction from sparse data is $(\lambda, \beta) = (255 \pm 4^\circ, -59 \pm 5^\circ)$ in J2000.0 ecliptic coordinates, which is not far from the value $(263^\circ, -67^\circ)$ derived by \cite{Dav.ea:96}. The difference between these two poles is $9^\circ$ of arc, which is within a typical uncertainty of the pole direction expected from models based on sparse data. The period $P = 4.633631 \pm 0.000005$\,h is close to the value $4.633638 \pm 0.000002 $\,h of \cite{Kaa.ea:01} based on a set of 40 lightcurves from 1988 to 1993. Fig.~\ref{fig:Ida_shapes} also illustrates a typical accuracy of the shapes derived from sparse photometry: they are only a rough approximation of the real (in general unknown) shape and cannot provide any surface details, but rather only global characteristics. They can be further refined with dense lightcurves or other complementary data. The relative accuracy of the period determination is however high, depending mainly on the length of the interval of observations. A typical uncertainty of the pole direction is 10--20$^\circ$, depending mainly on the number of brightness measurements, their photometric accuracy, and the number of apparitions.

\section{Distributed computing}

Because the problem of finding the global minimum in the parameter space consists of running the gradient optimization many times with different initial values and each run is independent of the others, the task is ideal for parallelization -- it is a so called embarrassingly parallel problem. This approach of scanning a range of periods is safe in the sense that we are sure that the best-fitting model is indeed the global minimum in the parameter space, because the convergence in the pole direction and shape is robust. 
An example of another approach to finding the global minimum, which will be implemented in the Gaia processing software \citep{Cel.ea:09}, is the use of genetic algorithms.

The default interval of periods we scan is usually 2--100 hours. The lower limit roughly corresponds to the spin barrier of asteroids larger than $\sim$200\,m \citep{Pra.ea:02b}, the upper limit was chosen arbitrarily -- the fraction of asteroids rotating so slowly is small and they are often in the excited rotation state that cannot be described by our simple model. These limits can be changed if needed. The time needed to scan the period interval for one asteroid with hundreds of measurements covering ten years or more is of the order of weeks at 1 CPU. It means that the processing of data for hundreds of thousands of known asteroids requires huge computational power. Therefore we set up a distributed computing project Asteroids@home\footnote{\url{http://asteroidsathome.net}} that uses idle time of tens of thousands computers of volunteers.

\section{BOINC}

The project is built on the Berkeley Open Infrastructure for Network Computing (BOINC) framework, which is major infrastructure for volunteer computing based on the SETI@home project \citep{Kor:12}. The BOINC server not only distributes unit tasks, collects and validates results, but also offers community-based tools like forums, credits system, user of the day acknowledgment, etc. The principles of BOINC and the parts of BOINC server and client are described in \cite{Kor:12} or \cite{Vin.Thi:13}. Here we cover only the setup specific for our project.

The period search interval for each asteroid is divided into few hundred smaller intervals. The number of intervals for a given asteroid depends on the number of data points, the length of observations, and the resolution of the shape model. It is set up such that the CPU time needed to process one interval is constant. Each computational unit contains initial values for the optimization, interval of periods, and the data. The data consist of time-brightness pairs and the geometry of observation: vectors towards Sun and the observer centered to the asteroid. Typically, the size of the input data file is less than 100\,kB. At volunteer's side, the lightcurve inversion gradient-based optimization is run. The $\chi^2$ values for each trial period are stored and sent back to the server. The typical size of the file sent back to the server is about 1\,MB or less. Each unit is sent to at least two volunteers and the results are compared and validated. The limit for delivering the unit is set to 14 days, an average time needed for finishing one unit is about three hours but can vary a lot across devices (hundreds of hours on mobile devices, for example). After finishing all units for a given asteroid, the periodogram is analyzed outside the BOINC platform and the best-fitting model and its uniqueness is checked. Although the best-fitting period and the corresponding shape model always exist, it is only rarely unique/prominent because of the low photometric quality of the currently available data.

To date (August 2015) $\sim$60,000 volunteers with $\sim$100,000 computers contributed to the project. The number of volunteers steadily increases with hundreds of new users each day. The current computing power is $\sim$150\,TFLOPs. The application is compiled for all main operating systems: 32- or 64-bit Windows and Linux, Mac OS, and FreeBSD. It can run also on mobile devices with Android and Raspberry Pi. A version for GPUs can run on CUDA 5.5 or higher. The code was optimized for speed, to further improve its performance, versions using SSE2/3 and AVX instructions sets are available.

\section{Data and Results}

As the source of photometric data, we currently use the so called Lowell Photometric Database \citep{Bow.ea:14}. It corresponds to the photometry reported to the Minor Planet Center (MPC), but the systematic shifts between the observatories were removed. Typically, the data set for one asteroid consists of hundreds of brightness measurements in $V$ band covering several years. The accuracy of the photometry is about 0.15-0.20\,mag. It means that in most cases, the signal is drowned by the level of random noise and systematic errors and there are many indistinguishable best-fitting solutions of the inverse problem. 

So far, we have processed about 100,000 asteroids out of 330,000 in the Lowell database. The preliminary results are posted on the web page of the project,\footnote{\url{http://asteroidsathome.net/scientific_results.html}} where the derived rotation period, the pole orientation, and an image of the shape model are shown. An important feedback for the volunteers is also the information about who was ``lucky enough'' to process the period interval that turned out to contain the global minimum. Because each unit is computed at least two times and then validated, there are two or more users assigned to each result. After further validation of reliability of solutions and their comparison with independent data -- periods from the Asteroid Lightcurve Database \cite{War.ea:09}, for example -- the models will be published in a peer-review journal and stored in the DAMIT database together with the data, shape files, and all other relevant information.

Although we have already processed tens of thousands of asteroids, the number of successfully derived models is low. This is because the quality of the data and their number is low, thus no unique period could be derived. A typical periodogram then does not have one or two clear minima like in the case of Ida in Fig.~\ref{fig:Ida_per}, but has many periods and corresponding models that fit the data to the same $\chi^2$ level. In such cases, only additional photometric data can help to derive a unique model.

\section{Future}

Processing all data from the Lowell Photometric Database is the first step on the way to reconstruct the distribution of physical properties of the asteroid population. Because the success rate is very low with these data, a crucial part of interpreting the results will be the proper de-biasing. For this, we will run an extensive test with synthetic data produced by realistic asteroid shapes and scattering and random distribution of spins and periods. We also plan to combine Lowell photometry with dense lightcurves available at the Minor Planet Center Asteroid Light Curve Database,\footnote{\url{http://www.minorplanetcenter.net/light_curve2/light_curve.php}} with future Gaia photometry, and with any photometry from other future surveys. Currently we are also computing shape models from all available photometry for several tens of asteroids with future mass estimates based on Gaia astrometry. Another challenge will be the combined inversion of photometric and thermal infrared data \citep{Dur.ea:12}.

\subsection*{Acknowledgments}

The success of Asteroids@home project would not be possible without tens of thousand of volunteers who provided computing resources of their computers. We greatly appreciate their contribution and patience. We also appreciate the help and support of enthusiastic members of the Czech National Team, who tested the application and significantly contributed to the optimization of the code. The work of J\v{D} was supported by the research grant 15-04816S of the Czech Science Foundation. JH is supported by the Centre national d'\' etudes spatiales (CNES) postdoctoral fellowship.

\newcommand{\SortNoop}[1]{}


\end{document}